# Design of compositionally graded contact layers for MOCVD grown high Al-content AlGaN transistors

Towhidur Razzak,[1,a)] Seongmo Hwang,[2] Antwon Coleman,[2] Hao Xue,[1] Shahadat Hasan Sohel,[1] Sanyam Bajaj,[1] Yuewei Zhang,[1] Wu Lu,[1] Asif Khan,[2] Siddharth Rajan[1]

[1] Department of Electrical and Computer Engineering, The Ohio State University, Columbus, Ohio, 43210, USA

[2] Department of Electrical Engineering, University of South Carolina, 301 Main Street, Columbia SC 29208, USA

In this letter, we design and demonstrate an improved MOCVD grown reverse Al-composition graded contact layer to achieve low resistance contact to MOCVD grown ultra-wide bandgap (UWBG) $Al_{0.70}Ga_{0.30}N$ channel metal-semiconductor field-effect transistors (MESFETs). Increasing the thickness of the reverse graded layer was found to improve contact layer resistance significantly, leading to contact resistance of $3.3 \times 10^{-5}$ $\Omega \cdot cm^2$. Devices with gate length, $L_G$, of 0.6 µm and source-drain spacing, $L_{SD}$, of 1.5 µm displayed a maximum current density, $I_{DS,MAX}$, of 635 mA/mm with an applied gate voltage, $V_{GS}$, of +2 V. Breakdown measurements on transistors with gate to drain spacing, $L_{GD}$, of 770 nm had breakdown voltage greater than 220 , corresponding to minimum breakdown field of 2.86 MV/cm. This work provides a framework for the design of low resistance contacts to MOCVD grown high Al-content $Al_xGa_{1-x}N$ channel transistors.

---

[a)] Author to whom correspondence should be addressed.  Electronic mail:  razzak.1@osu.edu

High Al-composition $Al_xGa_{1-x}N$ with $x \geq 0.7$ and bandgap, $E_G \geq 5.1$ eV are estimated to have high critical breakdown field, $F_{BR}$, and high saturated velocity, $v_{sat}$, which are attractive material properties for high voltage and high frequency power applications. This set of materials have $F_{BR}$ exceeding 11 MV/cm and correspondingly high lateral figure of merit, (LFOM)

$$LFOM = n_s \mu F_{BR}^2,$$

where $n_s$ is the sheet charge density in the channel and $\mu$ is the carrier mobility.[1] Since the saturated velocity $v_{sat}$ in $Al_xGa_{1-x}N$ is expected to be similar to GaN, the Johnson figure of merit (JFOM)

$$JFOM = \frac{v_{sat}F_{BR}}{2\pi},$$

for these materials are expected to be much higher than GaN.[2-4] Devices fabricated with high Al-composition $Al_xGa_{1-x}N$ will thus be more suitable for scaling due to the high $F_{BR}$. Compared to GaN, higher breakdown voltages are expected for devices with similar dimensions while maintaining comparable saturated current density, $I_{DS,MAX}$, due to similar $v_{sat}$.

However, establishing low-resistance Ohmic contact to $Al_xGa_{1-x}N$ with $x \geq 0.7$ is a challenge due to the low electron affinity in these materials. Although metal-semiconductor Ohmic contacts have been demonstrated for $x > 0.5$, the lowest contact resistance achieved to date is 24.6 Ω·mm.[5-10] On the other hand, reverse Al-composition graded contact layers have been very successful in reducing contact resistance down to 0.3 Ω·mm for molecular-beam epitaxy (MBE) grown $Al_{0.75}Ga_{0.25}N$ channel transistors and can be readily employed to form low resistance contacts, to any $Al_xGa_{1-x}N$ channel devices.[11] In these structures, the composition of $Al_xGa_{1-x}N$ in the contact region is gradually varied from the channel composition to a lower composition, while the material is doped with [Si$^+$] to ensure it remains n-type. The metal-semiconductor contact resistance is low in these cases since contact is being made to GaN. Since the material is gradually graded, there are no abrupt semiconductor band offsets to impede electron transport. These molecular beam epitaxy (MBE) grown devices, however, have significantly lower mobility than theoretically predicted.[12] The origin of the low mobility is still being investigated.[13] In contrast, metalorganic chemical vapor deposition (MOCVD) grown high Al-composition $Al_xGa_{1-x}N$ channels have been shown to have much higher mobility due to superior compositional uniformity, with electron mobility exceeding 150 cm$^2$/V·s.[5] Therefore, in this work we focus on achieving low-resistance MOCVD-grown graded contact layers to MOCVD-grown high Al-composition Al$_x$Ga$_{1-x}$N channels.

We now discuss some considerations necessary for the design of the reverse-graded contact layer. The (Al)GaN material system displays strong polarization effect due to the large piezoelectric polarization originating from the large ionicity of the (Al)Ga-N bonds and the strong spontaneous polarization resulting from the uniaxial nature of the crystal coupled with the non-ideal c/a ratio of the wurtzite structure.[14,15] Compositional grading of such polar materials produces bound polarization "bulk" charge density given by $D_\pi = -\nabla \cdot \mathbf{P}$, where $\mathbf{P}$ is the sum of spontaneous and piezoelectric polarization in Al$_x$Ga$_{1-x}$N alloys.[15] When the Al-composition in Al$_x$Ga$_{1-x}$N is graded from higher to lower Al-composition, a negative polarization charge is formed, causing a positive curvature in the energy band profile.[11] To compensate for this negative polarization charge and obtain a flat conduction band profile, it is necessary to introduce heavy n-type doping via [Si$^+$] incorporation in the lattice to produce a low resistance effective n-type region.[11] The resistivity of the compositionally graded contact layer, $\rho_{graded}$, can be described by

$$\rho_{graded} = \int_0^{t_{graded}} \frac{dz}{q \cdot \mu \cdot (N_D(z) - \nabla_z P[x(z)])} \quad , \quad (1)$$

where $q$, $\mu$, $N_D(z)$ and $P[x(z)]$ are the elementary electronic charge, carrier mobility, impurity doping density in the film and the polarization in the graded layer, respectively.

Doping of high Al-composition Al$_x$Ga$_{1-x}$N films has been studied for both MBE and MOCVD growth techniques. Although reports of conductive high Al-composition Al$_x$Ga$_{1-x}$N films exist for both methods, a discrepancy in the maximum achievable doping density has been observed between these two techniques.[16-21] While studies with MBE grown high Al-content Al$_x$Ga$_{1-x}$N films have consistently achieved high doping density above mid-10$^{19}$ cm$^{-3}$ for x < 0.8, such reports have been inconsistent for MOCVD grown samples.[16-21] The likely reason for this discrepancy is that for MOCVD grown Al$_x$Ga$_{1-x}$N films, dopant incorporation has been found to be a strong function of the Al-composition of the film and high dopant incorporation becomes more challenging for higher Al-content Al$_x$Ga$_{1-x}$N – which compounds the difficulty of growing reverse-graded contact layers for high Al-composition Al$_x$Ga$_{1-x}$N.[19,20] This is different from MBE-grown Al$_x$Ga$_{1-x}$N films where dopant incorporation was found to be independent of the Al-composition of the film.[16,17] Thus, while MBE can achieve degenerately doped Al-composition graded contact layer using the same growth condition, a continuous shift in optimum dopant incorporation condition makes growing such uniformly high doped Al-composition graded $Al_xGa_{1-x}N$ films challenging for MOCVD. Equation 1 predicts that for a 50 nm linearly down-graded contact layer with x graded from 0.7 to 0, like previous reports, an activated dopant density needs to be higher than $8\times10^{18}$ cm$^{-3}$, throughout the contact layer, to achieve low-resistance films – a challenging problem for MOCVD due to the reasons mentioned above.[22,23] This indicates that the contact layer resistance is sensitive to the dopant incorporation, and it is critical for the activated dopant density to be higher than the negative polarization charge density. Thus, to achieve lower contact resistance in MOCVD grown graded contact layers, it is necessary to reduce the negative polarization charge density in the graded contact layers, $\rho_{graded}$, by reducing the factor, $\nabla_z P[x(z)]$ – gradient of the polarization charge.

Two different contact layer structures were investigated for this study – both grown on an n-type $Al_{0.70}Ga_{0.30}N$ channel with a doping of $4.5\times10^{18}$ cm$^{-3}$. For Sample A (Figure 1(a)), the contact layer was graded from x=0.7 to x=0 over 50 nm, like previous reports, which requires [Si$^+$] > $8\times10^{18}$ cm$^{-3}$ for polarization charge compensation (Figure 1(b)).[22,23] For sample B (Figure 1(c)), the Al-composition was graded from x=0.7 to x=0.3 over 150 nm such that a lower [Si$^+$] concentration, can compensate the negative polarization charge density (Figure 1(d)). While a lower x at the top surface of the contact layer yields a lower metal-semiconductor resistance, the contact layer resistance will be higher due to a higher negative polarization charge density due to larger compositional grading and vice versa. A terminating Al-composition of x=0.3 was thus chosen for the second case, since metal-semiconductor contact resistance of low-$10^{-5}$ $\Omega$-cm$^2$ is expected for $Al_{0.3}Ga_{0.7}N$, which is sufficient for a proof of concept demonstration.[24] The thickness was chosen to be 150 nm, to reduce the negative polarization charge density even further so as to ensure a low resistance contact layer with a uniform $N_D = 3\times10^{18}$ cm$^{-3}$, which can be easily achieved for MOCVD for x < 0.8.[23]

The structures were grown using MOCVD on an AlN on sapphire template with a 100 nm channel with a doping density of $4.5\times10^{18}$ cm$^{-3}$ in the channel and $1\times10^{19}$ cm$^{-3}$ in the graded contact layer. The source/drain contacts were formed by depositing Ti/Al/Ni/Au metal stack using e-beam evaporation. The contacts on sample B were then annealed at 850 ºC for 30 seconds using a rapid thermal annealing (RTA) system to form alloyed contacts, followed by ICP-RIE plasma etch to define device isolation mesas. The active areas of the devices were defined by selectively recessing the graded contact layer between the source and drain contacts. Over-recessing was performed to ensure complete removal of the graded contact layer such that 40 nm of the channel remained after the contact-layer recess.

Electrical characteristics were measured using an Agilent B1500A semiconductor device analyzer. Two-terminal IV on the fabricated samples showed that sample A had non-linear IV characteristics which indicates the negative polarization charge in the contact layer has not been fully compensated. This indicates that the activated dopant density was lower than $8\times10^{18}$ cm$^{-3}$. The slow-graded contact scheme on sample B, however, shows linear IV characteristics, which indicates that the contact layer in sample B has indeed been compensated – as expected from the reduced polarization charge scheme. A comparison of two-terminal IV characteristics of the two samples are shown Figure 2. Hall measurements were performed on ungated four-terminal Van der Pauw (VDP) structures and the sheet resistance of the channel was found to be 5.6 k$\Omega/\square$ with sheet carrier density of $1.8\times10^{13}$ cm$^{-2}$ and mobility of 56 cm$^2$/V·s. Transfer length measurements (TLM) performed on gated TLM structures with recessed contact layer on sample B yielded a specific contact resistivity of $3.3\times10^{-5}$ $\Omega\cdot$cm$^2$. TLM and Hall measurements were not performed on sample A, due to the non-linear nature of the contacts. A comparison of the specific contact resistivity for state-of-the-art MOCVD-grown Al$_x$Ga$_{1-x}$N channel transistors with x > 0.5 is shown in Figure 3. As can be seen the specific contact resistivity obtained for sample B is the lowest observed for any MOCVD grown Al$_x$Ga$_{1-x}$N channel devices to date for x > 0.5. The contact layer resistivity was estimated to be approximately $1\times10^{-5}$ $\Omega\cdot$cm$^2$ which is, however, higher than MBE grown contact layers. Since Equation (1) predicts that the incorporated dopant density should be sufficient to compensate for the negative polarization charge and provide low resistance contact, the higher resistance could be originating from a) non-uniform grading of the contact layer leading to higher localized polarization charge and/or b) lower localized [Si$^+$] incorporation – both of which can lead to high resistance regions. This indicates that further growth optimization is required.

Metal-semiconductor field effect transistor (MESFET) structures were fabricated on sample B by depositing a gate metal stack of Ni/Au/Ni using e-beam evaporation. A schematic of the final fully processed sample structure is shown in Figure 4. Transfer IV characteristics (Figure 5(a)) measured for devices with L$_G$ = 0.6 µm and L$_{SD}$ = 1.5 µm (V$_{DS}$ = +20 V) showed a pinch-off voltage of -16 V and a maximum transconductance of 38 mS/mm. Output electrical characteristics (Figure 5(b)) measured on the same device showed a maximum current density of 635 mA/mm ($V_{GS}$ = +2 V), which to the best of our knowledge is the highest current density achieved for $Al_xGa_{1-x}N$ channel devices with x > 0.5 to date. 2-D technology computer aided design (TCAD) simulator, Silvaco, was used to model the described device structure.[25] As can be seen, good match is obtained for transfer characteristics and the saturated current density. However, some deviation is observed in the slope of the output curves in the linear region – most likely originating from a mismatch between the field dependent mobility model in Silvaco and the actual device characteristics. Three-terminal breakdown characteristics of the MESFETs (Figure 6) measured at $V_{GS}$ = −20 V for devices with gate-drain spacing, $L_{GD}$ = 0.77 µm, showed no breakdown up to $V_{GD}$ = +220 V, which translates to an average field of 2.86 MV/cm, almost 3× higher than that exhibited by lateral GaN channel devices with similar dimensions. The breakdown is mainly limited by the gate leakage current which is the primary contributor to the drain current in the three terminal breakdown measurement. Thus, the breakdown characteristics can be further improved by the addition of a gate dielectric such as $Al_2O_3$.[22]

In summary, we have designed and demonstrated an improved MOCVD grown Al-composition graded contact layer with a contact resistance of $3.3 \times 10^{-5}$ $\Omega \cdot cm^2$ – lowest for MOCVD grown $Al_xGa_{1-x}N$ channel transistor with x > 0.5. $Al_{0.7}Ga_{0.3}N$ channel transistors ($L_G$ = 0.6 µm and $L_{SD}$ = 1.5 µm) with this improved Al-composition graded contact layer exhibited a maximum current density of 635 mA/mm. Transistors with $L_{GD}$ = 0.77 µm did not undergo breakdown up to $V_{GD}$ = +220 V which corresponds to an average field of 2.86 MV/cm which is close to 3× higher than lateral GaN channel devices with similar dimensions. This demonstration provides a technologically important approach to form low resistance contacts to MOCVD grown UWBG $Al_xGa_{1-x}N$ channel transistors.

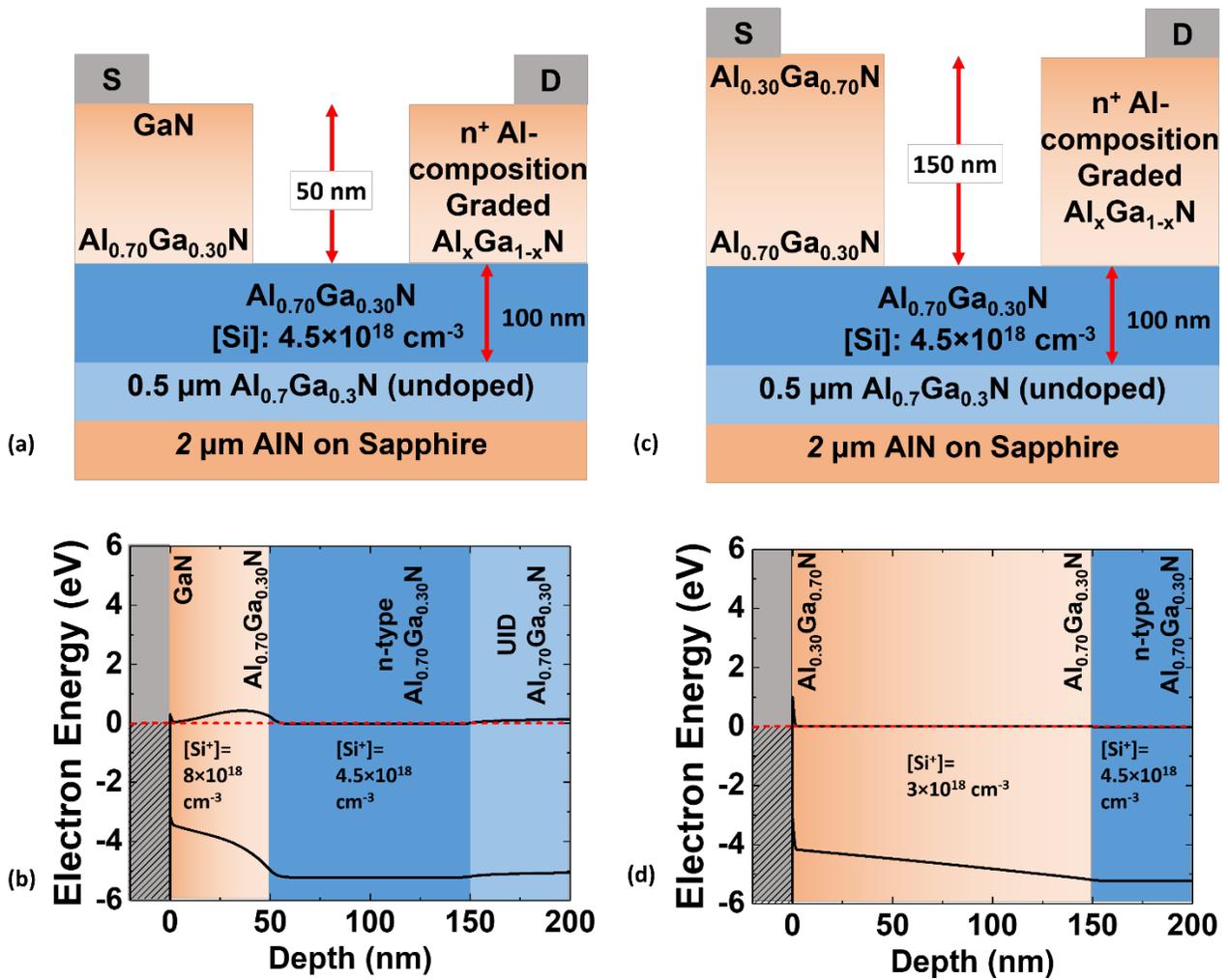

FIGURE 1. (a) Schematic and (b) energy band-diagram of the <u>access region</u> of sample A with 50 nm thick linearly reverse-graded contact layer with x graded from 0.7 to 0 and **[Si⁺] = 8×10¹⁸ cm⁻³** and (c) Schematic and (d) energy band-diagram of the <u>access region</u> of sample B with 150 nm thick linearly reverse-graded contact layer with x graded from 0.7 to 0.3 and **[Si⁺] = 3×10¹⁸ cm⁻³**

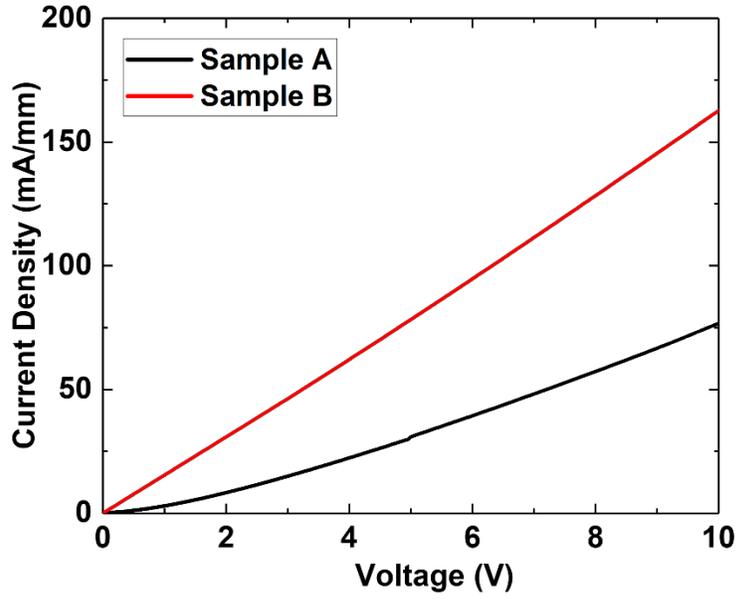

FIGURE 2. Comparison of two-terminal IV of sample A and B for two-terminal structures with $L_{SD}$ = 9 µm

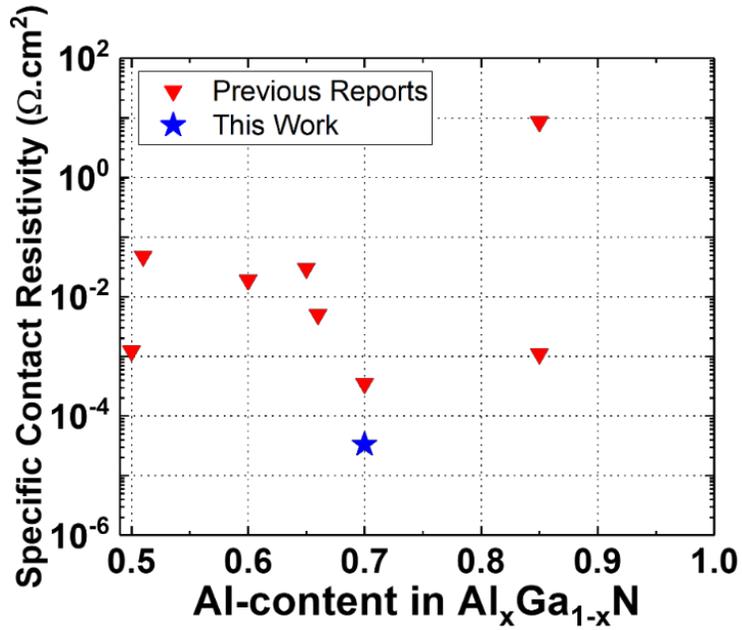

FIGURE 3. Comparison of the specific contact resistivity vs Al-content for state-of-the-art MOCVD-grown $Al_xGa_{1-x}N$ channel transistors with x > 0.5. [5-7,9,10,22,23,26,27]

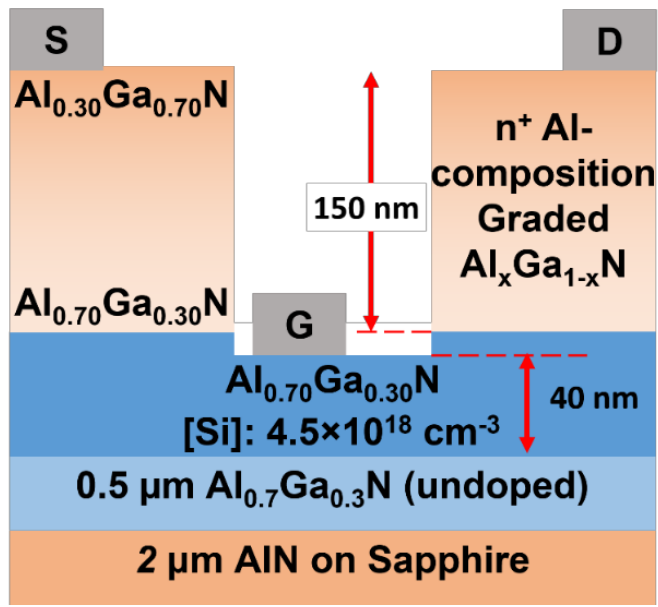

FIGURE 4. Schematic of the fully processed sample B with a 150 nm reverse Al-composition graded contact layer with a channel thickness of 40 nm

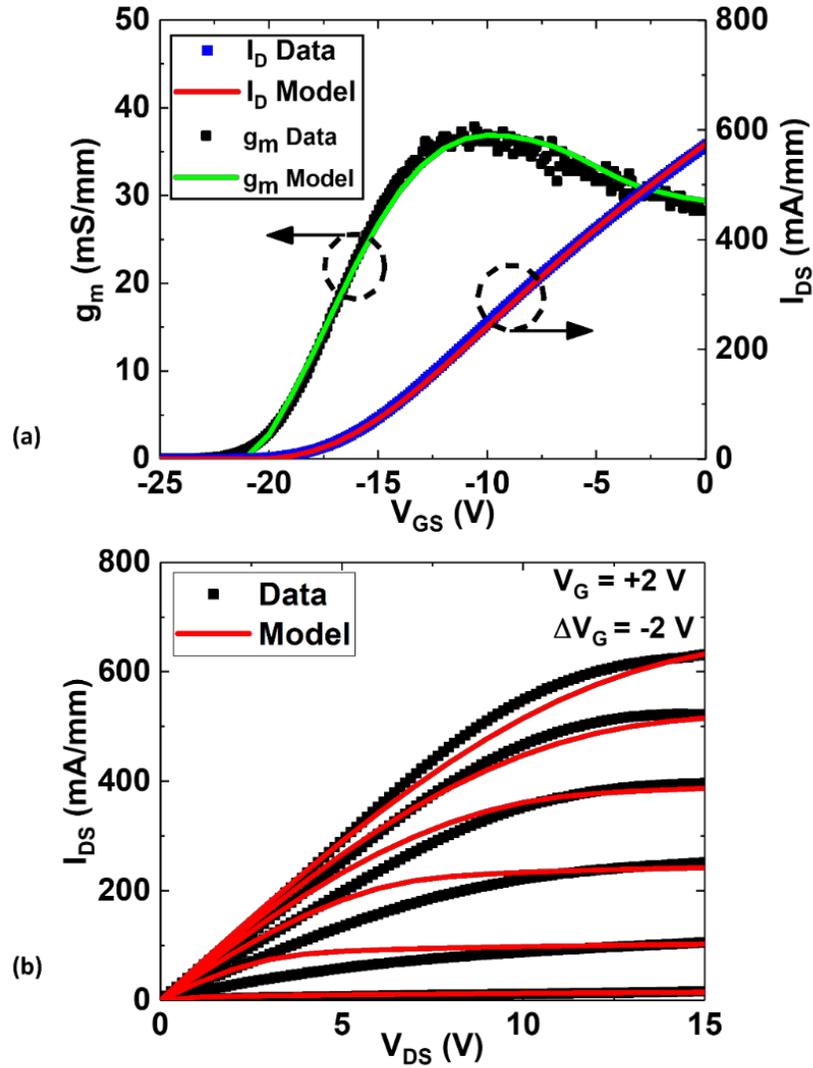

FIGURE 5. (a) Transfer characteristics and (b) family output I–V characteristics for device with gate-length, $L_G$ = 0.6 µm and source-drain spacing, $L_{SD}$ = 1.5 µm. Square symbols represent measured electrical characteristics while solid lines represent simulated characteristics.

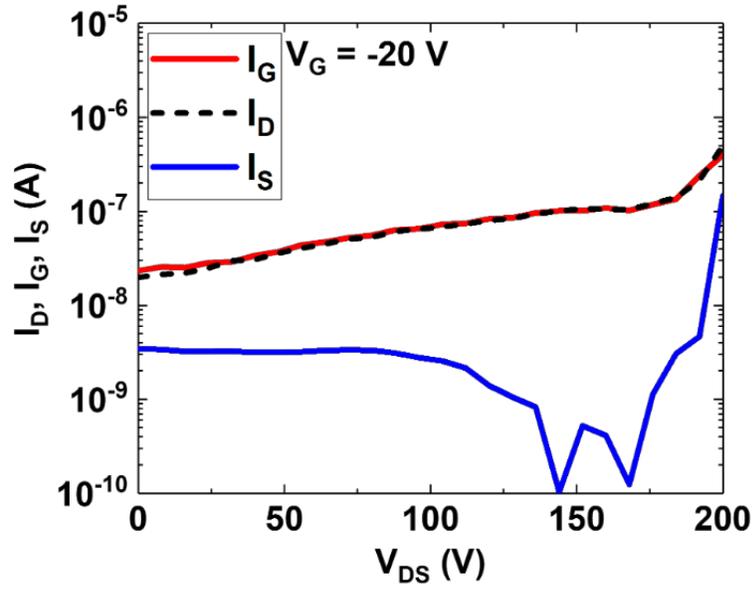

FIGURE 6. Three-terminal breakdown voltage measurement showed no breakdown up to $V_{GD}$ = +220 V measured at $V_{GS}$ = -20 V for a gate-drain spacing, $L_{GD}$ = 0.77 µm

**ACKNOWLEDGMENTS**

The authors acknowledge funding from Air Force Office of Scientific Research (AFOSR Grant FA9550-17-1-0227, Program Manager Kenneth Goretta) and the DARPA DREaM program (ONR N00014-18-1-2033, Program Manager Dr. Young-Kai Chen, monitored by Office of Naval Research, Program Manager Dr. Paul Maki).

**REFERENCES**

[1]   Jerry L Hudgins, Grigory S Simin, Enrico Santi, and M Asif Khan,  IEEE Transactions on Power Electronics **18** (3), 907 (2003).

[2]   AFM Anwar, Shangli Wu, and Richard T Webster,  IEEE Transactions on Electron devices **48** (3), 567 (2001).

[3]   Maziar Farahmand, Carlo Garetto, Enrico Bellotti, Kevin F Brennan, Michele Goano, Enrico Ghillino, Giovani Ghione, John D Albrecht, and P Paul Ruden,  IEEE Transactions on electron devices **48** (3), 535 (2001).

[4]   Towhidur Razzak, Hao Xue, Zhanbo Xia, Seongmo Hwang, Asif Khan, Wu Lu, and Siddharth Rajan, 2018 IEEE MTT-S International Microwave Workshop Series on Advanced Materials and Processes for RF and THz Applications (IMWS-AMP), 2018.

[5]   Andrew M Armstrong, Brianna A Klein, Albert Colon, Andrew A Allerman, Erica A Douglas, Albert G Baca, Torben R Fortune, Vincent M Abate, Sanyam Bajaj, and Siddharth Rajan,  Japanese Journal of Applied Physics **57** (7), 074103 (2018).

[6]   Albert G Baca, Andrew M Armstrong, Andrew A Allerman, Erica A Douglas, Carlos A Sanchez, Michael P King, Michael E Coltrin, Torben R Fortune, and Robert J Kaplar,  Applied Physics Letters **109** (3), 033509 (2016).

[7]   Sakib Muhtadi, S Hwang, Antwon Coleman, Fatima Asif, A Lunev, MVS Chandrashekhar, and Asif Khan,  Applied Physics Letters **110** (17), 171104 (2017).

[8]   Sakib Muhtadi, Seong Mo Hwang, Antwon Coleman, Fatima Asif, and Asif Khan, Device Research Conference (DRC), 2017 75th Annual, 2017.

[9]   Sakib Muhtadi, Seong Mo Hwang, Antwon Coleman, Fatima Asif, Grigory Simin, MVS Chandrashekhar, and Asif Khan,  IEEE Electron Device Letters **38** (7), 914 (2017).

[10]  Norimasa Yafune, Shin Hashimoto, Katsushi Akita, Yoshiyuki Yamamoto, Hirokuni Tokuda, and Masaaki Kuzuhara,  Electronics Letters **50** (3), 211 (2014).

[11]  Sanyam Bajaj, Fatih Akyol, Sriram Krishnamoorthy, Yuewei Zhang, and Siddharth Rajan, Applied Physics Letters **109** (13), 133508 (2016).

[12]  Sanyam Bajaj, Ting-Hsiang Hung, Fatih Akyol, Digbijoy Nath, and Siddharth Rajan,  Applied Physics Letters **105** (26), 263503 (2014).

[13]  Sanyam Bajaj, The Ohio State University, 2018.

[14]  Siddharth Rajan, Huili Xing, Steve DenBaars, Umesh K Mishra, and Debdeep Jena,  Applied physics letters **84** (9), 1591 (2004).

[15]  Debdeep Jena, Sten Heikman, Daniel Green, Dario Buttari, Robert Coffie, Huili Xing, Stacia Keller, Steve DenBaars, James S Speck, and Umesh K Mishra,  Applied physics letters **81** (23), 4395 (2002).

[16]  Jeonghyun Hwang, William J Schaff, Lester F Eastman, Shawn T Bradley, Leonard J Brillson, David C Look, J Wu, Wladek Walukiewicz, Madalina Furis, and Alexander N Cartwright, Applied Physics Letters **81** (27), 5192 (2002).

[17]  B Borisov, V Kuryatkov, Yu Kudryavtsev, R Asomoza, S Nikishin, DY Song, M Holtz, and H Temkin,  Applied physics letters **87** (13), 132106 (2005).

18     Yoshitaka Taniyasu, Makoto Kasu, and Naoki Kobayashi, Applied physics letters **81** (7), 1255 (2002).

19     Pietro Pampili and Peter J Parbrook, Materials Science in Semiconductor Processing **62**, 180 (2017).

20     Frank Mehnke, Tim Wernicke, Harald Pingel, Christian Kuhn, Christoph Reich, Viola Kueller, Arne Knauer, Mickael Lapeyrade, Markus Weyers, and Michael Kneissl, Applied Physics Letters **103** (21), 212109 (2013).

21     Y-H Liang and E Towe, Applied Physics Reviews **5** (1), 011107 (2018).

22     Sanyam Bajaj, Andrew Allerman, Andrew Armstrong, Towhidur Razzak, Vishank Talesara, Wenyuan Sun, Shahadat H Sohel, Yuewei Zhang, Wu Lu, and Aaron R Arehart, IEEE Electron Device Letters **39** (2), 256 (2018).

23     Sakib Muhtadi, S Hwang, Antwon Coleman, Fatima Asif, A Lunev, MVS Chandrashekhar, and Asif Khan, Applied Physics Letters **110** (19), 193501 (2017).

24     Brianna Alexandra Klein, Albert G Baca, Andrew M Armstrong, Andrew A Allerman, Carlos Anthony Sanchez, Erica Ann Douglas, Mary H Crawford, Mary A Miller, Paul G Kotula, and Torben Ray Fortune, ECS Journal of Solid State Science and Technology **6** (11), S3067 (2017).

25     *Device Simulation Software, ATLAS User's Manual*. (Silvaco Int., Santa Clara, CA, 2009).

26     EA Douglas, S Reza, C Sanchez, D Koleske, A Allerman, B Klein, AM Armstrong, RJ Kaplar, and AG Baca, physica status solidi (a) **214** (8), 1600842 (2017).

27     Hirokuni Tokuda, Maiko Hatano, Norimasa Yafune, Shin Hashimoto, Katsushi Akita, Yoshiyuki Yamamoto, and Masaaki Kuzuhara, Applied Physics Express **3** (12), 121003 (2010).